\documentclass[preprint2]{aastex}
\usepackage{rotating}
\usepackage[T1]{fontenc}
\usepackage[latin9]{inputenc}
\usepackage{array}
\usepackage{float}
\usepackage{graphicx}
\usepackage{amssymb}
\usepackage{subfigure}
\usepackage{multirow}
\usepackage{longtable}
\usepackage{changepage}

\makeatletter

\begin{document}

\title{A lingering non-thermal component in the GRB prompt emission: predicting GeV emission from the MeV spectrum}

\author{Rupal Basak and A.R. Rao}

\affil{Tata Institute of Fundamental Research, Mumbai - 400005, India. $rupalb@tifr.res.in,
arrao@tifr.res.in$}

%\email{rupalb@tifr.res.in}
\begin{abstract}
The high energy GeV emission of gamma-ray bursts (GRBs), detected by \emph{Fermi}/LAT, has a significantly 
different morphology compared to the lower energy  MeV emission, detected by \emph{Fermi}/GBM. Though the late time GeV emission
is believed to be synchrotron radiation produced via an external shock, this emission as early as the prompt 
phase is puzzling. Meaningful connection between these two emissions can be drawn only by 
an accurate description of the prompt MeV spectrum. We perform a time-resolved spectroscopy of the GBM data 
of long GRBs having significant GeV emission, using a model consisting of 2 blackbodies and a power-law.
We examine in detail the evolution of the spectral components and found that GRBs having high GeV 
emission (GRB~090902B and GRB~090926A) have a delayed onset of the power-law component, in the GBM spectrum, which lingers 
at the later part of the prompt emission. This behaviour mimics the flux evolution in LAT. In contrast, 
bright GBM GRBs with an order of magnitude lower GeV emission (GRB~100724B and GRB~091003) show a coupled 
variability of the total and the power-law flux. Further, by analyzing the data for a set of 17 GRBs, we find a strong correlation 
between the power-law fluence in the MeV and the LAT fluence (Pearson correlation: r=0.88 
and Spearman correlation: $\rho=0.81$). We demonstrate that this correlation is not influenced 
by the correlation between the total and the power-law fluences at a confidence level of 2.3$\sigma$. We speculate the 
possible radiation mechanisms responsible for the correlation.

\end{abstract}


\section{INTRODUCTION}
Gamma ray burst (GRB) was first discovered in late 1960's as a flash of near MeV photons, known as the prompt phase. 
It took nearly a quarter of a century to observe the higher energy (GeV) photons. The first detection was
in the afterglow of GRB 940217 (Hurley et al. 1994), observed by \textit{CGRO}/EGRET, 90 minutes after the \textit{CGRO}/BATSE 
detection of the prompt emission. Later, it became apparent that the high energy emission is also present during the prompt 
phase either as a simple extrapolation of the prompt spectral model (Dingus et al. 1998) or as an additional spectral component
(Gonzalez et al. 2003). The origin of the high energy photons, however, remains speculative. For example, they could be produced by
internal/external shocks via leptonic or hadronic mechanism, and/or via magnetic jet (e.g., Meszaros \& Rees 1994, 2011;
Waxman 1997; Fan \& Piran 2008; Panaitescu 2008; Zhang \& Pe'er 2009). Though there is a 
rich structure predicted by theoretical models, these can be realised only by detectors having good spectral resolution and 
wide band coverage (Zhang et al. 2011).

With the advent of the Fermi satellite, we have a wider energy coverage with unprecedented sensitivity.
The Fermi satellite hosts two instruments --- the Gamma-ray Burst Monitor (GBM), a dedicated instrument for GRB detection, 
and the Large Area Telescope (LAT). GBM covers 8 keV to 30 MeV (Meegan et al. 2009), while LAT covers 20 MeV to 300 GeV
(Atwood et al. 2009). Recently, Ackermann et al. (2013; A13 hereafter) have released the first LAT GRB catalog, which contains a total of 35 
GRBs (also see Akerlof et al. 2011; Rubtsov et al. 2012). In order to find possible association between the LAT and GBM emissions in GRBs,
they have studied the fluence in GBM and LAT in ``GBM'' time window (see their Figure 17). LAT fluence is
calculated independently by GBM-LAT joint fit and LAT-only analysis. For brighter bursts they have found disagreements due to
multiple components in the GBM-LAT joint analysis. Since the high energy emission generally lasts longer, they have performed the
same study in ``LAT'' time window to account for the correct energetics of LAT. This set contains 19 GRBs (17 long GRBs).
Though they found tentative trend of GBM-LAT correlation, the data scatter is high, and more importantly, they have found 
two sets of GRBs: hyper-fluent LAT bursts (080916C, 090510, 090902B, and 090926A) and the rest. The LAT photons can be detected 
during or outside the prompt emission time window. Hence, to get an uniformity of data, Zheng et al. (2012; Z12 hereafter) selected a sample 
of 22 GRBs (17 long GRBs), restricting the time window for match filter technique to 47.5 s interval following the associated GBM trigger. 
They found a rather poor correlation --- Pearson correlation coefficient of 0.537.

This lack of a strong association between the MeV and the GeV emission could be due to
the spectral diversity in the prompt emission. Zhang et al. (2011) have made
a joint analysis of the time resolved spectra across the full band of GBM and LAT detectors and
have identified 5 possible combination of spectral models (e.g., Band --- Band et al. 1993, blackbody+power-law etc.).
One of the limitations of such time-resolved analysis, however, is the limited statistics
available for finer time bins: a smaller time bin for a time resolved analysis 
results in poor count statistics whereas a broad time bin will be unable to capture 
the spectral evolution adequately. Recently Basak \& Rao (2013, hereafter BR13) have 
assumed certain spectral evolution for a given spectral model to reduce the number of free parameters to describe the 
individual pulses of a GRB (also see Basak \& Rao 2012b). 
%{\bf Rupal: The following details is moved to a later place - please check}
%For example, if there are 10 time bins, a four parameter model (e.g., Band) requires 40 
%parameters to study the full pulse. However, certain parameters (e.g., photon indices --- %$\alpha$, $\beta$ of Band model)
%vary little over short duration (e.g., falling part of a pulse). Hence, the time resolved %spectra should
%be fitted joinly, tying these parameters at all times. Also, some parameters have well %behaved variation,
%e.g., peak energy ($E_{peak}$) of Band model falls off at the falling part of a pulse. %Hence, the evolution can be 
%parametrized. With these assumptions BR13 could successfully reduce the number of free %parameters, e.g., from
%40 parameters, they could describe a pulse by just 14 parameters. 
BR13 assumed various spectral models
(for example, Band, blackbody with a power-law --- BBPL, multicolour blackbody with a power-law --- mBBPL,
two blackbodies with a power-law --- 2BBPL) and performed joint parametrized fit.
They have shown that the 2BBPL model is superior to a single blackbody with a power-law (BBPL) for the individual pulses 
of two GRBs, namely 
GRB 081221 and GRB 090618. Moreover, the 2BBPL model shows marginal superiority ($\sim$ 70\% confidence) to the Band model in some cases. 
Though the physical origin of the 2BBPL model
is only speculative at this moment, it has some attractive features, e.g., the temperature and the normalization of the two BBs are
highly correlated. In fact, BR13 put these constraints on the 2BBPL model, still they always found better $\chi^2_{red}$
than BBPL, mBBPPL and Band model. 

It has been shown for BATSE data (Ryde 2004) and GBM data (Ryde et al. 2010, Zhang et al. 2011) that the model 
consisting of a thermal and non-thermal component have comparable or sometimes statistically better fit than the Band model 
in the initial bins. Further, for BATSE data it has been shown that the power-law component becomes progressively 
important at the later part (Gonzalez et al. 2003). Remembering that the GeV emission has a delayed onset, it can be speculated 
that the power-law component in the prompt emission drives the GeV photons.
Since a time-resolved joint fit to the MeV and GeV data could not identify unique spectral
models (Zhang et al. 2011),  
in this work, we investigate the possibility of making a parametrized-joint fit to
the MeV data and identifying spectral components in them which can be
used to predict the
%are correlated to the
%GeV a joint fand address the question of predictability of
LAT fluence. The plan of the paper is as follows. We  discuss the data selection and analysis method in Section 2.
Results are discussed in Section 3. In Section 4, we  draw our conclusions and discuss some issues.

\section{Data selection and Analysis}
The A13 catalog has 17 long GRBs for GBM-LAT fluence study. 5 GRBs in this set has 
either much delayed onset than LAT or only upper limit on GBM fluence. Z12 set ignores the following GRBs: 090323, 090328, 
090626, 091031 and 100116A, and takes additional 5 GRBs, namely, 091208B, 100325A, 100724B, 110709A and 120107A. As we are interested 
in the connection between the GBM and LAT during the prompt emission, we need uniform time selection, and hence we use Z12 set of long 
GRBs for a correlation study.

We closely follow the parametrized-joint fit technique, devised by BR13.
Since our attempt is to segregate the prompt MeV spectrum into thermal (blackbodies) and
non-thermal (power-law) parts to test whether we can predict the LAT fluence, we choose
2BBL as the preferred spectral model. In some of the GRBs, we
verified that indeed the 2BBPL model is preferred over the other models. For example,
in three episodes of GRB~090902B, the $\chi^2$ (dof) of Band, BBPL, mBBPL and 2BBPL are as follows. 
Episode 1 (0.0 to 7.2 s): 1066.0 (894), 1221.8 (888), 973.9 (886), 983.5 (886). Episode 2
(7.2 to 12.0 s): 5778.6 (1515), 1876.4 (1501), 1731.9 (1500), 1735.2 (1499). Episode 3
(12.0 to 35.2 s): 4181.0 (3137), 5142.4 (3108), 3853.5 (3107), 3796.9 (3106). We note that the 2BBPL
model is much better than the Band model, in all episodes. The only comparable model is mBBPL, but 2BBPL 
is still better than this model in episode 3, which in fact covers 2/3rd of the duration. Moreover, in
BR13, it was  found that 2BBPL is better than mBBPL in all cases.
  
In the following we give a brief description of the methodology for 2BBPL model fit. 
2BBPL model has the following parameters --- temperatures ($kT_{\rm 1}$, $kT_{\rm 2}$) and normalizations
($N_{\rm 1}$, $N_{\rm 2}$) of the two BBs, and power-law index ($\Gamma$) and normalization ($N_{\rm \Gamma}$) 
of the power-law component. We found that the temperature and normalizations of the 2 BBs are highly correlated
($kT_{\rm 2}=x. kT_{\rm 1}$ and $N_{\rm 2}=y.N_{\rm 1}$). We use this relation in all time bins while fitting in XSPEC. 
We take all the parameters as free for the first bin. 
For all other bins, e.g., $\rm i^{ th}$ bin, kT$_{\rm 1}$(i) and N$_{\rm 1}$(i) are free, while 
$kT_{\rm 2}\rm (i)=kT_{\rm 1}\rm(i)\times \frac{kT_{\rm 2}\rm (1st bin)}{kT_{\rm 1}\rm (1st bin)}$,
$N_{\rm 2}\rm (i)=N_{\rm 1}\rm(i)\times \frac{N_{\rm 2}\rm (1st bin)}{N_{\rm 1}\rm (1st bin)}$.
It does not matter in which bin we choose all the parameters free, XSPEC determines the most appropriate ratio `x' and `y' to minimize 
the $\chi^{\rm 2}$. For the power-law component, we assume that the index can be tied in all bins.
Note that we have dropped the parametrization scheme of BR13,
as the current GRBs are not well structured as broad separable pulses.  

The time bins for the spectral fits are chosen by requiring equal number of counts in each time bin. This minimum count is
chosen  between 800-1200 taking into account the peak count and duration. Only for 3 cases, namely, GRB 090902B, GRB 090926A and 
GRB 100724B, which have the highest GBM fluence, we take the minimum count to be 2000, 2000, and 1800 respectively. 
For GRB 081006, which has very low GBM count, we could use only one bin from -0.26 to 5.9 seconds (see A13). 
The spectra are then binned as described by
BR13 --- i.e., for NaI detectors, one bin in 8-15 keV, 7 or 5 bins in 100-900 keV, with progressively higher bin size at higher energies,
and for BGO detectors, 5 bins in 200 keV-30 MeV, with progressively higher bin size at higher energies. 
For example, spectral rebinning reduces 128 channels of NaI of Fermi/GBM to $\sim$50 bins. If we demand
20 counts per channel, this requires 1000 counts per time bin, which is roughly the requirement of the time cuts that we have put.
We calculate the fluxes 
of each model component, in each time bin. We propagate the normalization errors to calculate the errors in fluxes. These are then used
to calculate the fluence, with the corresponding error for the individual components, and the total model. We use the LAT event count, provided by Z12.
Note that the LAT fluence are calculated in the 47.5 s time window. 
We calculate the fluence quantities of GBM both in the $T_{90}$ (provided by A13) 
and within the time window of 47.5 s.

To study the correlation between different 
fluence values we use the Pearson and Spearman rank correlation. The associated chance probabilities are also calculated.
To determine which of the correlations is more fundamental we use the Spearman partial rank correlation method (Macklin 1982). This method
enables one to analyse the correlation between two variables, say A and X, in the presence of another variable, say Y. The significance
level associated with the correlation between A and X, independent of Y is given by a D-parameter, which gives in terms of $\sigma$, the confidence level at which it
can be stated that the correlation between A and X is not influenced by Y.
%is a function of total number
%of data points and the partial Spearman rank correlation of A and X, given Y. 
To fit the scattered data, we use the 
a linear model of the form log(y) = K + $\delta$log(x), using the technique of joint likelihood for the coefficients K and $\delta$
(D'Agostini 2005; Basak \& Rao 2012a). Following D'Agostini (2005), we put a gaussian `noise' parameter ($\sigma_{int}$), denoting 
the intrinsic scatter of the data in the y-coordinate. This formalism is useful if y depends on extra `hidden' variables.
%, which we do not `see'.

\section{Results}
\begin{table*}\centering
%\begin{adjustwidth}{-1.6cm}{}
%\small
%\hspace{-1.1in}
 \caption{The set of 17 GRBs and their properties}
%\begin{small}
 \hspace{-1.5in}

 \begin{tabular}{c|c|cc|cc|c}

\hline
 GRB & Count & \multicolumn{2}{c|}{GBM $T_{90}$ window$^{(a)}$} & \multicolumn{2}{c|}{47.5 s time window } & LAT fluence\\
 & for time-cut & \multicolumn{2}{c|}{(Photon cm$^{-2}$)} & \multicolumn{2}{c|}{(Photon cm$^{-2}$)} & (Photon m$^{-2}$)\\

\cline{3-6}
& & Total fluence & PL fluence & Total fluence & PL fluence &  in 47.5 s\\
\hline
\hline 
080825C & 1200 & 224.8$\pm$6.2 & 105.2$\pm$5.5 & 245.1$\pm$13.3 & 115.7$\pm$11.9 & 36.6$\pm$11.6\\
080916C & 1200 & 369.9$\pm$7.7 & 223.9$\pm$6.4 & 329.3$\pm$5.7 & 196.34$\pm$4.8 & 279.0$\pm$24.9\\
081006A$^{(b)}$ & --- & 6.97$\pm$0.92 & 3.24$\pm$0.62 & 6.97$\pm$0.92 & 3.24$\pm$0.62 & 16.3$\pm$4.7\\
090217 & 1000 & 124.7$\pm$4.1 & 54.1$\pm$3.4 & 129.0$\pm$4.4 & 55.8$\pm$3.7 & 22.5$\pm$6.0\\
090902B & 2000 & 1028.4$\pm$18.6 & 498.3$\pm$14.6 & 1102.7$\pm$29.6 & 525.3$\pm$23.2 & 378.1$\pm$29.5\\
090926A & 2000 & 739.6$\pm$10.8 & 324.9$\pm$8.6 & 785.6$\pm$13.1 & 343.3$\pm$10.4 & 372.2$\pm$28.0\\
091003 & 1000 & 186.8$\pm$6.3 & 95.9$\pm$4.6 & 210.1$\pm$9.8 & 107.7$\pm$7.2 & 14.8$\pm$4.5\\
091208B & 800 & 60.5$\pm$3.5 & 37.2$\pm$3.1 & 82.5$\pm$11.4 & 43.6$\pm$10.0 & 14.6$\pm$6.5\\
100325A & 1200 & 13.4$\pm$1.7 & 3.3$\pm$0.9 & 13.9$\pm$1.7 & 3.6$\pm$0.9 & 6.7$\pm$3.0\\
100414A & 1200 & 289.9$\pm$7.6 & 103.4$\pm$6.2 & 384.8$\pm$7.9 & 145.4$\pm$6.4 & 87.5$\pm$33.1\\
100724B & 1800 & 998.5$\pm$9.5 & 500.6$\pm$7.1 & 396.6$\pm$3.8 & 212.4$\pm$2.9 & 23.9$\pm$7.6\\
110120A & 1000 & 69.1$\pm$4.7 & 27.5$\pm$2.2 & 77.9$\pm$7.2 & 32.6$\pm$3.4 & 9.5$\pm$3.6\\
110428A & 800 & 127.4$\pm$3.5 & 32.4$\pm$2.6 & 147.6$\pm$5.5 & 44.0$\pm$4.0 & 8.0$\pm$3.6\\
110709A & 1000 & 198.9$\pm$5.6 & 92.2$\pm$5.2 & 212.1$\pm$6.2 & 101.1$\pm$5.7 & 18.7$\pm$7.1\\
110721A & 1200 & 182.2$\pm$7.2 & 98.8$\pm$4.5 & 192.5$\pm$9.6 & 105.0$\pm$6.0 & 46.4$\pm$9.3\\
110731A & 1000 & 89.6$\pm$5.7 & 55.5$\pm$2.0 & 102.9$\pm$11.9 & 66.9$\pm$4.1 & 81.5$\pm$10.4\\
120107A & 800 & 39.5$\pm$4.1 & 25.8$\pm$4.9 & 39.7$\pm$5.2 & 25.8$\pm$3.9 & 17.6$\pm$7.2\\

\hline
\end{tabular}

\label{list}
%\end{small}

\begin{footnotesize} 

$^{(a)}$ $T_{90}$ values are taken from A13\\
$^{(b)}$ $T_{90}$ value is retained for larger window\\
%$^b$ $\rm T_{90}$ is taken from vander Horst (2008). The average flux over -1.0 to 8.0 s is used for $\rm T_{90}$ as well. 
\end{footnotesize}

%\end{adjustwidth}
\end{table*}

\subsection{The lingering non-thermal component}

\begin{figure*}\centering
{

\includegraphics[width=6.5in]{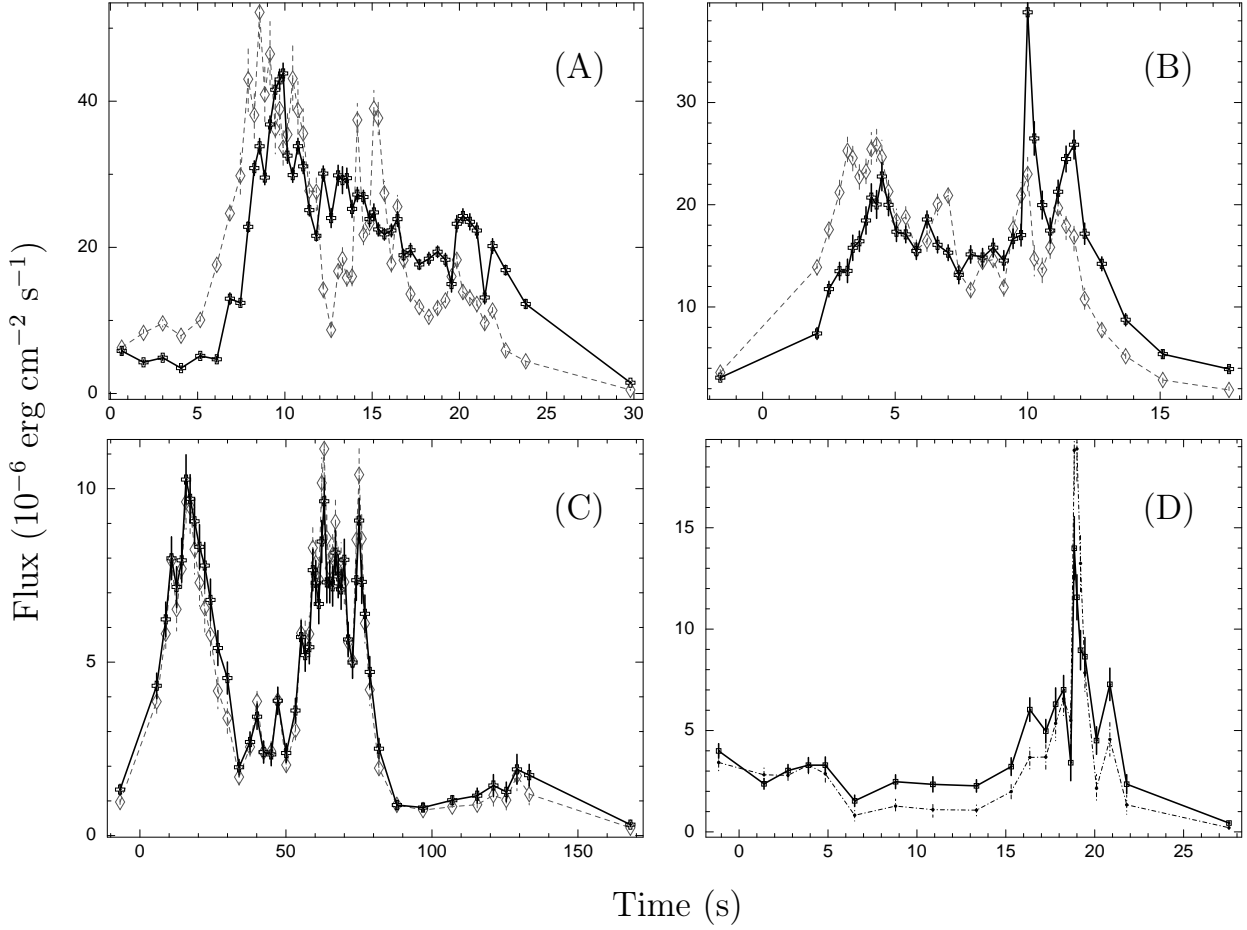} 

%\label{fig:Light Curves}
}
\caption{Total (dashed line) and non-thermal (solid line) components in various GRBs. The non-thermal flux is multiplied by a constant, 
namely, the ratio of average total and non-thermal flux to plot them on the same scale. \textit{(upper panels):} cases where the 
non-thermal component has delayed onset and persists at the later part. LAT count for these are high --- (A) 378.1 for GRB 090902B, 
(B) 372.2 for GRB 090926A. \textit{(lower panels):} cases where the non-thermal component tracks the thermal component and LAT count 
is low --- (C) 23.9 for GRB 100724B, (D) 14.8 for GRB 091003.
}
\label{linger}
\end{figure*}

Figure 5 of Z12 shows the scatter plot between the LAT photon counts and GBM photon counts. In this figure, we can see 
that for similar GBM fluences the LAT photon count can vary by more than an order of magnitude.
We identify two pairs of GRBs: pair 1 contains GRB 090902B and GRB 090926A; the other pair contains GRB 100724B and GRB 091003.
These pairs, despite having comparable fluence in GBM, have widely different LAT fluence. Note that the GRBs in pair 1 have the 
highest fluence among the hyper-fluent LAT GRB class (A13), which contains 4 GRBs (3 long). As described in
Section 2, we segregate the thermal and non-thermal part and analyze the GBM data by following BR13. 
Note that by `thermal' we mean the two blackbodies, which may or may not have thermal orgin. On the other hand,
we consider the power-law as the `non-thermal' component. 
In Figure~\ref{linger}, we show the energy flux evolution of the total 
and the non-thermal components for the indvidual GRBs. The upper panels show the flux evolution of the first pair and the lower panels 
show that of the second pair. It is clear that there is a delayed onset of the non-thermal component for GRB 090902B and GRB 090926A. 
This component
dominates at the later part of the prompt emission. This behaviour was first reported by Gonzalez et al. (2003) for GRB 941017.
Note that the LAT fluence over 47.5 s of these GRBs are quite high --- 378.1 and 372.2 photon m$^{-2}$, respectively. 
On the other hand, the non-thermal and the total flux of GRB 100724B and GRB 091003 originates almost at the same time and 
their flux evolution more or less tracks each other. The LAT fluence are 23.9 and 14.8 photon m$^{-2}$, respectively. Hence, 
it seems that there is indeed a strong morphological difference between GRBs having high and low LAT counts.
We make comparison with LAT light curves of the corresponding GRBs in A13, and find that
the PL component of the GBM data, independent of the LAT data, mimics the LAT behaviour.
\subsection{MeV-GeV correlations}
\begin{table*}\centering
 \caption{Correlations between the LAT fluence with the GBM fluence and GBM PL fluence }
 \begin{tabular}{c|c|c|c|c|c}
\hline
 Correlation & \multicolumn{2}{c|}{Pearson} & \multicolumn{3}{c}{Spearman} \\
\cline{2-6}

       & r &  $\rm P_{r}$ & $\rho$ &  $\rm P_{\rho}$ & D\\
\hline
\hline 
Ia$^{(a)}$ & 0.68 & $2.67\times10^{-7}$ & 0.73 & $8.20\times10^{-4}$ & -0.6\\
IIa & 0.68 & $2.67\times10^{-7}$ & 0.79 & $1.66\times10^{-4}$ & 1.8\\
\hline
Ib & 0.87 & $5.66\times10^{-6}$ & 0.75 & $5.61\times10^{-4}$ & -1.4\\
IIb & 0.88 & $3.20\times10^{-6}$ & 0.81 & $9.23\times10^{-5}$ & 2.3\\

\hline
\end{tabular}
\label{corr}
\begin{footnotesize} 

Note: 
$^{a}$ See text for detail
\end{footnotesize}

\end{table*}

We study two kinds of correlations: (I) GBM-LAT fluence correlation and (II) non-thermal GBM fluence-LAT fluence correlation.
If the GBM fluence is measured in $T_{90}$ then we call it `a', and if the fluence is measured in 47.5 s time bin we call it `b'.
In Table~\ref{list}, we list the various fluence quantities of the GRBs. 
The LAT photon fluences are quoted from Z12 in the last column.

\begin{figure*}\centering
{

\includegraphics[width=6.5in]{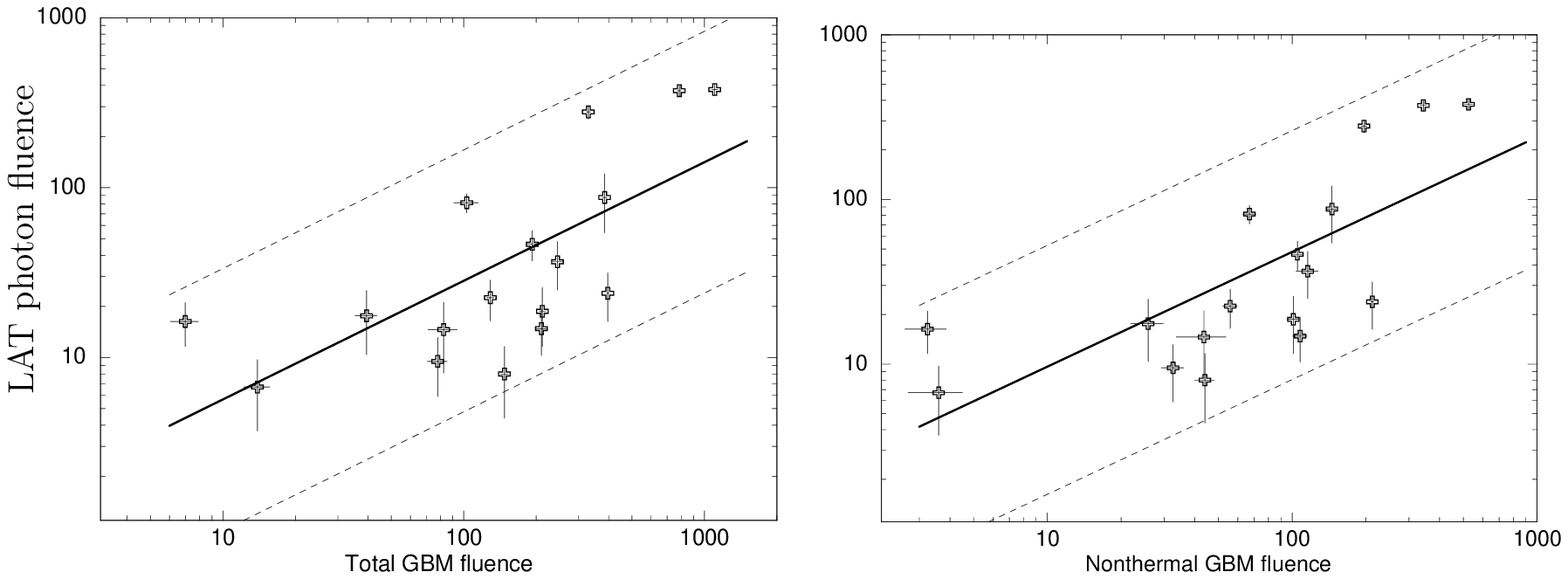} 

%\label{fig:Light Curves}
}
\caption{Correlations between GBM fluence (photons cm$^{-2}$) with LAT fluence (photons m$^{-2}$), both calculated within 
the time window 47.5 s.
}
\label{correlation}
\end{figure*}

In Figure~\ref{correlation}, we give a scatter plot of Ib and IIb, respectively, as described above. In Table~\ref{corr}, we 
report the correlation coefficients of these plots. The p values denote the chance probability of these correlations. Hence, lower this
value the better is the confidence of the correlation. Note that the Pearson correlation of IIb is marginally better than Ib. As Pearson
correlation is unable to determine which among Ib and IIb is more fundamental, we use the Spearman partial correlation test. 
Note that the Spearman correlation is a more robust estimator of a correlation 
(Macklin 1982) as it does not depend on the linearity of the data. Also, the correlation is least affected by outliers. We note that the
Spearman correlation ($\rho$) of Ib and IIb are 0.75 and 0.81, respectively. The D-parameter, which denotes the significance of the 
correlation between two variables, in presence of a third parameter, is shown in the last column. Note that the value is negative for 
correlation Ib, denoting that this correlation is affected by the correlation between GBM fluence and GBM non-thermal fluence. The D-value
of correlation IIb, on the other hand, is 2.3, denoting that this correlation is more fundamental at a significance of 2.3$\sigma$, while
there is a correlation between GBM fluence and GBM non-thermal fluence. Similar inferences can be drawn if we use $\rm T_{90}$ instead of 
47.5 s interval (compare Ia with IIa in this case). Note that the GBM-LAT fluence correlation of Z12 is 0.537, while we 
get a correlation of 0.68. This may be due to different values of $T_{90}$ and the spectral models. Also note that we have calculated the Pearson correlation 
of the actual data. If the logarithmic values are used, we get the following correlations: 0.65 for Ia, 0.69 for Ib 0.68 for IIa and 0.72 for IIb. 

In order to find the relation between the GBM fluence and LAT fluence, we fit the scattered data of correlation Ib and IIb,
as described in Section 2. The results of the linear fits are shown in Table~\ref{fit}. K, $\delta$ are y-intercept and slope of the straight
line, respectively. Note that $\sigma_{int}$ is lower in case IIb, denoting that we have better knowledge about this correlation.
In Figure~\ref{correlation}, we have shown the fits by solid lines. The dashed lines denote the 2$\sigma_{int}$ scatter of the data.

\begin{table*}\centering
 \caption{Linear fit results of the scattered plots of Figure 2}
 \begin{tabular}{c|c|c|c|c}
\hline
 Correlation & K & $\delta$ & $\sigma_{int}$ & $\chi^2_{red}$ (dof)\\
\hline
\hline 
GBM-LAT (Ib) & $0.056\pm0.099$ & $0.698\pm0.044$ & $0.385\pm0.084$ & 1.10 (15) \\
GBM PL-LAT (IIb) & $0.288\pm0.096$ & $0.697\pm0.049$ & $0.368\pm0.081$ & 1.10 (15) \\

\hline
\end{tabular}
\label{fit}

\end{table*}

\section{Discussion and  Conclusions}

The origin of the GeV emission in GRB is still an open question. It is essential to study the 
GeV emission in order to understand the prompt emission and the afterglow. The power-law decay 
of the late GeV emission suggests that the emission might be synchrotron radiation produced 
via external shock (e.g., Kumar \& Barniol Duran 2010), when the fireball runs into the external 
medium. However, the production of GeV photons as early as the prompt emission itself is unexplained. 
Attempts have been made to use the MeV-GeV data to fit a model for the full energy band (e.g., Abdo et al 2009).
These schemes have failed to connect the prompt MeV-GeV emission in a global sense --- (a) there is no 
unified spectral model which can explain the full energy range (e.g., Zhang et al. 2011 have found 5 
combinations of them, and more importantly, models other than Band is required for high 
count cases) (b) the correlation between MeV and GeV is too weak to draw inferences (Z12). 

Meaningful connection can be drawn only by an accurate description of the prompt spectrum
and its evolution. The fact that spectral evolution during the prompt phase is not arbitrary and behaves 
smoothly with time gives us a better handle on the data. Using this technique for various models, BR13 
have shown that 2BBPL is the best compared to other popular models, most 
notably the Band model. This is the motivation of using this model for the present analysis.

To check the predictive power of this new model, we applied this technique to the GBM data of
the set of 17 GRBs. The idea was to check the morphology of various components in the GBM data 
alone and predict the LAT data. We found that prediction of GeV emission is possible if we segregate 
the model components. We found, for the first time, that the power-law of our model, despite 
the fact that the data is only GBM data, mimics the behaviour of the LAT data. More importantly, 
the total fluence of this component has a very strong correlation with the LAT fluence. 
This is a very exciting result as we have a prediction of prompt GeV data from the MeV data itself. 

Gupta \& Zhang (2007) and Fan \& Piran (2008) have considered various possibile radiation mechanisms 
which can lead to the GeV emission. If we consider internal shock model (e.g., Rees \& Meszaros 1994) for 
MeV emission and extrapolate the spectrum to GeV, then it 
over-predicts the GeV emission. Le \& Dermer 
(2009) showed that the low detection rate of LAT is consistent by assuming a ratio of 0.1 between GeV 
and MeV emission (but also see Guetta, Piran \& Waxman 2011). Beniamini et al. (2011) have considered
a set of 18 GRBs, having the highest luminosity in GBM and still undetected in LAT. They have obtained 
an average upper limit of LAT/GBM fluence ratio of 0.13 (in the prompt phase) and 0.45 (in the 600 s 
time window). These ratios put strong constraints on the prompt-afterglow models and particularly rule 
out synchrotron self compton model (SSC, e.g., see Meszaros, Rees \& Papathanassiou 1994; 
Fan \& Piran 2008) for both MeV and GeV emission.

The implication of our finding is that the GeV emission is not driven by the full
prompt MeV emission, but the power-law component drives it. This puts more stringent
constraints on the different models of GeV afterglow emission. For example, consider SSC 
as the mechanism of late GeV emission. The fact that the prompt emission flux
is shared by the 2BB and power-law, and that only the power-law drives the GeV emission means 
we need to put only the energy of this power-law component in the calculation. This, for example,
decreases the limit of the highest possible circumburst density (n) than the standard model (Wang et al. 2013). 
For GRB 090902B, which has low reported n, our condition makes the SSC implausible for this case. 

In summary, though the GeV emission has a significantly distinct morphology than the MeV emission, 
they are connected through a component of the prompt MeV emission itself --- the power-law shares the common origin 
with the prompt GeV emission.

\section*{Acknowledgments} This research has made use of data obtained through the
HEASARC Online Service, provided by the NASA/GSFC, in support of NASA High Energy
Astrophysics Programs.

\end{document}